\documentclass[5p]{elsarticle}
\usepackage{epsfig}
\usepackage{a4wide}
\usepackage{amssymb}
\usepackage{amsmath}
\usepackage{graphicx}
\usepackage{float}
\usepackage{microtype}

\newcommand{\be}{\begin{equation}}
\newcommand{\ee}{\end{equation}}
\newcommand{\ba}{\begin{eqnarray}}
\newcommand{\ea}{\end{eqnarray}}

\begin{document}\sloppy

\vspace{\baselineskip}

\title{
{%
\vspace{-3.0cm}
\small\hfill\parbox{64.0mm}{\raggedleft%
TTP13-006, SFB/CPP-13-09
}}\\[0.5cm]
 On the ${\cal O}(\alpha_s^2)$ corrections to 
$b \to X_u e \bar \nu$ 
inclusive decays }

\author{Mathias Brucherseifer}
\address{ Institut f\"ur Theoretische Teilchenphysik, 
Karlsruhe Institute of Technology (KIT), Karlsruhe, Germany}

\author{Fabrizio Caola and Kirill Melnikov}

\address{
Department of Physics and Astronomy,
Johns Hopkins University,
Baltimore, MD, USA}

\begin{abstract}
\noindent We present  
${\cal O}(\alpha_s^2)$  QCD corrections to 
the fully-differential decay rate of a $b$-quark 
into inclusive semileptonic charmless final states. 
Our calculation  provides  genuine two-loop QCD corrections, 
beyond the Brodsky-Lepage-Mackenzie (BLM)  approximation,  
to {\it any } infra-red safe partonic observable that can be probed 
in $b \to X_u e \bar \nu$  decays. Kinematic cuts 
that closely match those 
used in experiments can be fully accounted for.  To illustrate these points, 
we compute  the non-BLM corrections  to moments of the hadronic invariant 
mass and the hadronic energy with  cuts on the lepton 
energy and the hadronic invariant mass. 
Our results remove one of the sources of theoretical uncertainty
that affect 
the extraction of the CKM matrix element $|V_{ub}|$ from charmless 
inclusive $B$-decays.
\end{abstract}

\maketitle

Studies of CP violation in $B$-mesons performed by BELLE and BABAR, firmly 
established the correctness of  the Cabbibo-Kobayashi-Maskawa paradigm at 
the few percent level. These studies will be continued,  when the super-$B$ 
factory in Japan will come on line.
A powerful tool to test the CKM picture is the unitarity of 
the CKM-matrix. A combined  fit to all available data gives 
$|V_{ub}| = 3.58(13) \times 10^{-3}$ \cite{ufit}. This number should 
be compared with the value $|V_{ub}| = 3.38(36) \times 10^{-3}$ extracted  
from exclusive $B \to h e \bar \nu$ 
decays, where the hadron $h$ is either a pion or a $\rho$-meson, 
and with  $|V_{ub}| = 4.27(38) \times 10^{-3}$ which is obtained from 
{\it inclusive} measurements of $B \to X_u e \bar \nu$ decays 
\cite{pdg}. Although exclusive and inclusive  results 
are not in serious disagreement, they are clearly  different and further 
scrutiny of both exclusive 
and inclusive determinations of $|V_{ub}|$ is certainly warranted. 

The most complicated theoretical issue for both exclusive and inclusive methods 
is the control of non-perturbative effects. This is hard to do for exclusive 
decays and important input in this case is provided by {\it ab initio} 
lattice 
QCD calculations of exclusive $B \to \pi,\rho,..$ transitions.  In contrast, 
in case of 
{\it inclusive}  semileptonic decays  of $B$-mesons,  non-perturbative 
difficulties 
can be largely circumvented by the application  of  local operator 
product expansion (OPE) \cite{ope1,ope2,ope3}.  
The OPE  allows to compute 
sufficiently inclusive  observables related to 
semileptonic decays of $B$-mesons, such as the total rate and moments 
of various kinematic distributions,  by correcting 
distributions and rates  of semileptonic decays of $b$-quarks  with a
limited number
of universal non-perturbative parameters. 
These non-perturbative parameters 
can be determined from fits to semileptonic decays of $B$-mesons 
to charmed final states $B \to X_c e \bar \nu$
\cite{fit1,fit1a,fit2,babar2009,Gambino:2004qm,Gambino:2011cq}
and  then used  in the description of 
 $B \to X_ue \bar \nu$ transitions,
facilitating the extraction of the CKM matrix element $|V_{ub}|$ 
from observables in the latter. 

While this procedure is well-defined  theoretically, it was not used 
in the determination of $|V_{ub}|$ right away because 
$B \to X_ue \bar \nu$ transitions suffer from a much larger 
$B \to X_ce \bar \nu$ background.
One can place  severe  cuts on the kinematics 
of final state particles to suppress it; for example,  
requiring  that the hadronic invariant mass is  smaller 
than the mass of the $D$-meson, $m_D \sim 1.87~{\rm GeV}$, clearly 
eliminates the charm background.    
However, it was realized early on that  such cuts  lead 
to problems with the convergence of the operator product 
expansion and infinitely many  terms in the OPE 
need to be summed up  to obtain reliable results.  Such a resummation is 
usually
expressed through the so-called shape function \cite{neub,bigi} which 
parametrizes the residual motion 
of a heavy quark inside a heavy meson.  A recent discussion 
of $B \to X_u e \bar \nu$ decay in the shape-function 
region, that includes next-to-next-to-leading order (NNLO) 
QCD effects, can be found  in Ref.~\cite{gnp}.
Unfortunately, current uncertainties
in the functional form of both leading and sub-leading shape-functions 
are significant and affect a precise determination of $|V_{ub}|$. 

In parallel to the studies of the shape function region, 
it was suggested that 
a  combination of cuts 
on  hadronic and leptonic invariant masses \cite{bauer} allows 
one to extend the phase-space coverage in $B \to X_u e \bar \nu$ 
decays and make the impact  of the shape functions 
smaller. Measurements  that use 
these selection criteria  
were  performed by the BELLE collaboration \cite{belle34}.
Further  advances in experimental techniques allowed to achieve 
an  almost  complete phase-space coverage in  $B \to X_u e \bar \nu$ 
decays.  Indeed, in recent experimental measurements it was possible 
to fully reconstruct
 the $B \bar B$ kinematics from their decay products, thereby   
allowing  to  extend selection cuts for the $b \to u$ process  
into the charm-rich regions and yet,  successfully reject 
the $b \to X_c e\bar \nu$ background.  
For example, two recent measurements by 
BELLE \cite{belle1} and  BABAR \cite{bab1} 
present  partial decay 
rates and a variety of kinematic distributions for $b \to u$ transitions 
with the cut on the electron energy as low as $E_{l} > 1~{\rm GeV}$.  
These cuts are inclusive enough so that the local OPE expansion 
can be used with confidence to describe $B \to X_u e \bar \nu$ decays. 

We summarize now  the status of the theoretical 
description of  $B \to X_u e \bar \nu $ 
decays,  under the assumption that the local OPE is applicable. 
The OPE expansion in the inverse $b$-quark 
mass   
$m_b$ is well-established for moments of the hadronic invariant mass 
and the hadronic  energy  \cite{ope1,ope2,ope3}.
The leading order term 
in the OPE 
expansion is given by the partonic $b \to u$ transition.  
The total decay rate for $b \to u$ is known in perturbative QCD through 
${\cal O}(\alpha_s^2)$ \cite{timo} 
and a large number of kinematic distributions and 
their moments are known through ${\cal O}(\alpha_s)$
\cite{ural1,jk,cz,li}. Also, the so-called BLM 
${\cal O}(\beta_0 \alpha_s^2)$ corrections \cite{Brodsky:1982gc}, 
that can be derived  by considering  the 
contribution of a massless $q \bar q$ pair 
to the $b \to u$ transition, 
are known 
for the  decay rate and main kinematic distributions
\cite{Luke:1994du,hoang,Gambino:2006wk}.
The only kinematic distribution in $b \to u$ decays 
that is known beyond the BLM approximation is the 
electron-neutrino
invariant mass 
distribution, 
computed in  Ref.~\cite{cz1}.
While the BLM-approximation is known to account for a significant 
fraction of the complete ${\cal O}(\alpha_s^2)$ correction, the precision 
of  current and, especially, forthcoming measurements of $|V_{ub}|$, 
the relatively large value of $\alpha_s(m_b)$ 
and a large variety of kinematic cuts employed in experimental analyses
make it very desirable to compute NNLO QCD corrections 
to the fully-differential $b \to u $ decay rate 
beyond the BLM  approximation.   The goal of this paper is to provide 
such a computation.

The calculation of NNLO QCD corrections to the $b \to u e \bar \nu $ 
decay requires three ingredients: {\it i}) two-loop amplitudes  for the
$b \to u e \bar \nu$  transition; {\it ii}) one-loop amplitudes for  
$b \to ug e\bar \nu$; {\it iii}) tree  amplitudes for 
$b \to u gge \bar \nu$ and $b \to u q \bar q e \bar \nu$.  
The two-loop amplitudes were computed 
by several authors in recent years \cite{bon,bell,astr,ben}. 
The one-loop amplitudes for $b \to uge\bar \nu$ can be extracted from 
the computation reported in Ref.~\cite{Campbell:2005bb}.  Finally, 
the tree   amplitudes for 
$b \to u gg e \bar \nu$ and $b \to u q \bar q e \bar \nu$ are straightforward 
to calculate  and compact results can be obtained by using the spinor-helicity 
formalism. These amplitudes  can be found in Ref.~\cite{bck}.

The well-known challenge for fully-differential NNLO QCD computations 
is to put these different contributions  together in a consistent way.
This is not easy to do since individual contributions 
 exhibit infra-red and collinear 
divergences and correspond to processes with different final-state 
multiplicities. For the computation reported in this paper, we use 
a method proposed in Refs.~\cite{Czakon:2010td,Czakon:2011ve} 
(see also  \cite{Boughezal:2011jf}) which combines the idea of 
sector decomposition  \cite{binothheinrich1,binothheinrich2,an1}  
with the phase-space partitioning \cite{Frixione:1995ms} in such a way 
that singularities are extracted from matrix elements  in 
a process-independent way. This framework leads to a parton level 
integrator which can be used to compute an arbitrary number of kinematic 
distributions in a single run of the program.  
We have recently given  a detailed description
of the relevant computational techniques in a paper ~\cite{bck} 
that describes a calculation  of NNLO QCD corrections to a related 
process $t \to be^+\nu$  and so we do not repeat it  here.  
Instead, we focus  on the illustration of 
phenomenological capabilities of the program that are relevant for the 
description of  $b \to u$ transitions. 

\begin{figure}[t]
\centering
\includegraphics[scale=0.55]{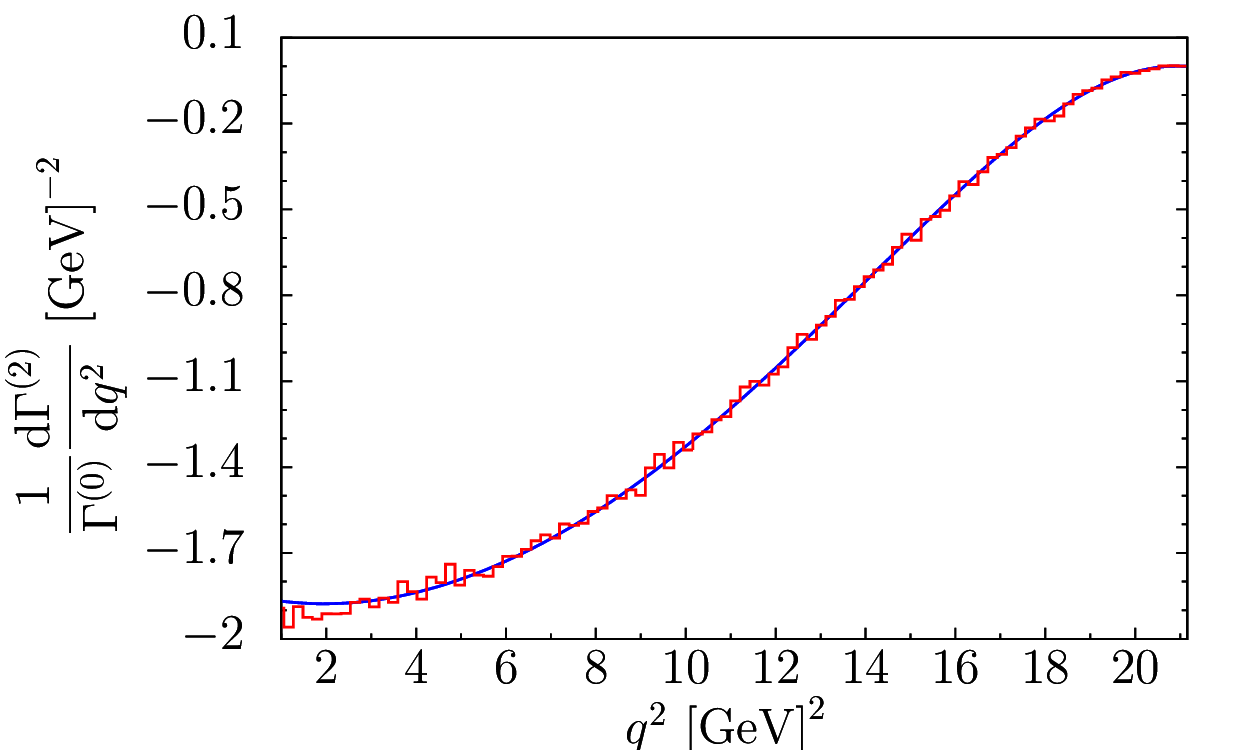}
\caption{
The coefficient of the second order correction to the lepton 
invariant mass distribution. The solid curve is the analytic result 
of Ref.~\cite{cz1}.
}\label{fig1}
\end{figure}

Numerical results reported below are obtained within the standard framework 
for perturbative QCD computations. We 
employ the on-shell renormalization 
for  the $b$-quark field and the $b$-quark mass. 
The strong coupling constant is  
renormalized  in the ${\overline {\rm MS}}$-scheme.
We note that we do not include the charm mass dependence when we compute 
contributions of additional $q \bar q$ pairs to the decay rate. 
As was explicitly shown in Ref.~\cite{bell}, contributions of  virtual 
charm loops for physical value of $m_c$ can be obtained, with a good 
accuracy,  from the bottom quark loops by equating charm and 
bottom masses. We will use this recipe in what follows. 
We write the differential decay 
rate for $b \to X_u e \bar \nu $ through NNLO in perturbative QCD as 
\be
{\rm d} \Gamma = {\rm d} \Gamma^{(0)} 
 + a_s {\rm d} \Gamma^{(1)} 
 + a_s^2 {\rm d} \Gamma^{(2)}+
{\cal O}(\alpha_s^3),
\ee
where $a_s = \alpha_s/\pi$ and 
$\alpha_s$ is  the ${\overline {\rm MS}}$
strong coupling constant at the scale $\mu = m_b$.
By integrating the fully differential decay rate over all the available
phase-space for final state particles, we obtain a prediction for the 
${\cal O}(\alpha_s^2)$ correction to the total decay rate. We write 
the result of our numerical integration in the following way
\be
\begin{split} 
 \Gamma^{(2)} =  \Gamma^{(0)} & \left ( 
 -29.98(8) + 2.143(7) N_f 
\right. \\
& \left.  - 0.0243 N_h \right ),
\end{split} 
\label{eq1} 
\ee
where $N_f=3$ denotes the number of massless quarks in the theory 
and $N_h=2$ denotes the number of quarks whose mass coincides 
with the $b$-quark mass.
Also,  $\Gamma_b^{(0)} = G_F^2 |V_{ub}|^2 m_b^5/(192 \pi^3)$ is the total 
decay rate for $b \to u e \bar \nu$ at leading order in perturbative 
QCD. Comparing  our computation to the analytic results presented 
in Ref.~\cite{timo}, we find agreement for each term shown  in Eq.(\ref{eq1})
to better than five per mille.

Having reproduced the known result for the NNLO QCD 
corrections to the total rate, we can now proceed to the discussion of 
kinematic distributions.  Our numerical program is set up in such a way 
that it can compute various kinematic distributions, both conventional
and cumulative, in a single run.  To illustrate this, we show 
in Fig.~\ref{fig1} ${\rm d} \Gamma^{(2)}/{\rm d} q^2$, where $q^2$ is 
the invariant mass of the lepton pair.  The solid curve is the 
result of the analytic calculation from  Ref.~\cite{cz1}. The numerical
and analytical results perfectly agree for all values of $q^2$ except 
in the region $q^2\sim 0$ where some discrepancy is observed.  This
discrepancy is not surprising since the 
analytic results of Ref.~\cite{cz1} were  obtained as 
an expansion around $q^2=m_b^2$ so that  deviations at small $q^2$ reflect 
convergence problems of the analytic computation in that
region.

\begin{figure}[t]
\centering
\includegraphics[scale=0.77]{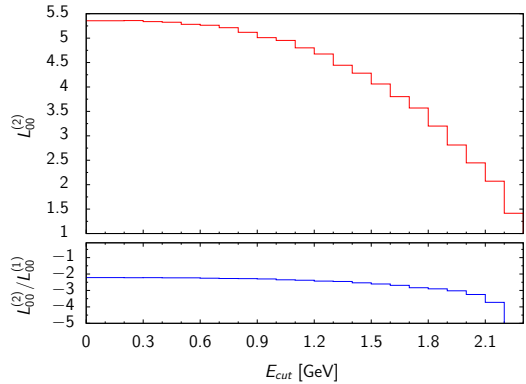}
\caption{
The cumulative histogram that shows $L^{(2)}_{00}$ 
as a function of the cut on the  charged lepton energy 
$E_l > E_{\rm cut}$. No  cut on the hadronic invariant 
mass is  applied.
}\label{fig2}
\end{figure}

To further discuss kinematic distributions, we 
follow Ref.~\cite{ural1} and  define 
moments of the partonic invariant mass $M_X^2=(p_b-p_e-p_\nu)^2$ 
and energy $E_X=E_b-E_e-E_\nu$, in dependence of the lower cut on the 
electron  energy $E_{\rm cut}$  
and the upper cut on the partonic invariant mass $M_{\rm cut}$. More 
specifically, we write
\be
L_{ij} = \langle M_X^{2 i}  E_X^{j} \theta(E_{e} - E_{\rm cut})
\theta(M_{\rm cut}  - M_{X} )
\rangle\label{defL}
\ee
where $\langle ... \rangle$ denotes the normalized 
phase-space average for final-state particles in 
$b \to X_u e \bar \nu$
\be
\langle \mathcal F \rangle \equiv \frac{1}{\Gamma^{(0)}} \int \mathrm{d} \Gamma \mathcal F.
\ee
We note~\cite{ural1} that one can use  $L_{ij}$'s 
defined in Eq.(\ref{defL}) to obtain 
moments of the lepton invariant mass $q^2$.

We write the moments in Eq.(\ref{defL}) 
as  an  expansion in the strong coupling 
constant and explicitly separate the BLM corrections 
\be
L_{ij} = 
L_{ij}^{(0)} + a_s  L_{ij}^{(1)}
+ a_s^2 \left ( 
 L_{ij}^{(2), \rm BLM}
+ L_{ij}^{(2)} \right ).
\ee
The BLM correction to these moments is 
obtained by computing the contributions of a massless $q \bar q$ pair 
$L_{ij}^{(2),n_f}$ 
and then by rescaling it by the full leading order QCD $\beta$-function
for three massless flavors 
\be
L_{ij}^{(2), \rm BLM} =  -27/2 \;L_{ij}^{(2),n_f=1}.
\ee

For the numerical calculation of the moments reported below, 
we use $m_b = 4.6~{\rm GeV}$ as the value of the $b$-quark pole mass.
To specify partonic 
cuts, we introduce the physical hadronic invariant mass 
\be
M_H^2   = {\bar \Lambda}^2 + 2 m_b \bar \Lambda E_X + m_b^2 M_X^2,
\label{eq5}
\ee
where ${\bar \Lambda} = m_{B^\pm} - m_b = 0.6769~{\rm GeV}$, for our choice 
of the $b$-quark mass.   We impose a cut on $M_H$ and  translate it 
into a cut on $M_X$ and $E_X$ using Eq.(\ref{eq5}).  Throughout 
the paper, we  use $M_H <  2.5~{\rm GeV}$ as the cut on the hadronic 
invariant mass.  Also, as we already mentioned, 
the BLM corrections are well-known.  They were 
discussed previously in the literature (see e.g.  Ref.~\cite{ural1})
and, for this reason, we focus on non-BLM corrections  in the remainder 
of this paper.

\begin{figure}[t]
\centering
\includegraphics[scale=0.77]{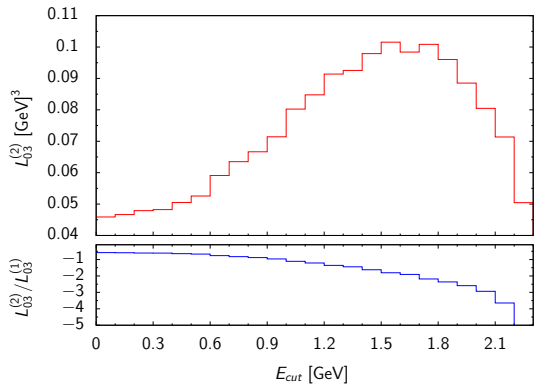}
\caption{
The cumulative histogram that shows $L^{(2)}_{03}$ 
as a function of the cut on the charged lepton  energy 
$E_l > E_{\rm cut}$. No  cut on the hadronic invariant 
mass is  applied. 
}\label{fig3}
\end{figure}

Our results for the 
moments are presented  
in Figs.~\ref{fig2},\ref{fig3} and in  
Table~\ref{table1}.
In Figs.~\ref{fig2},\ref{fig3}  we show  cumulative 
histograms for the non-BLM contributions to two moments 
$L_{0j}^{(2)}$ for $j=0$ and $j=3$, 
with no cut on  hadronic invariant mass. In both cases, the 
$x$-axis shows  the applied cut on the  the charged lepton energy. 
These figures illustrate  that our numerical program  
works as a parton level Monte Carlo integrator and that it can 
reliably compute large number of infra-red safe  
observables for the $b \to X_ue \bar \nu$ 
decay with  various  cuts in a {\it single 
run}.  This should be useful for further studies  of 
charmless decays of $B$-mesons  given, in particular,  a large 
number of   kinematic cuts  employed  in experimental 
analyses.\footnote{We also note that our numerical program is rather 
fast. For example, all numerical results reported in this paper, 
including distributions shown in Figs.~\ref{fig1},\ref{fig2},\ref{fig3} and in 
Table~\ref{table1}  were obtained in an overnight run on a modest-size  
computer cluster.}

To illustrate  the dependence of the non-BLM  corrections on the 
applied cuts,  we show the 
ratio of non-BLM contributions $L^{(2)}$ to NLO ones $L^{(1)}$,
as a function of the  electron  energy cut 
in lower panes of Figs.~\ref{fig2},\ref{fig3}. 
We note that this ratio is renormalization-scale independent.
We are interested in this ratio because, if 
it is  independent of $E_{\rm cut}$,  we could have found 
the corrections to the moments without fully-differential NNLO computations. 
However,  it is apparent from Figs.~\ref{fig2},\ref{fig3}
that this is not possible and that 
 non-BLM  corrections have a different functional dependence on 
$E_{\rm cut}$  as compared to the NLO ones. 
In addition, the cut-dependence is 
strongly moment-dependent  and it is more pronounced 
for higher-$j$ moments.  

We will now take a closer look at the numerical values of the computed 
corrections.   To facilitate this, we show in 
Table~\ref{table1} our results for leading, next-to-leading and 
next-to-next-to-leading order {\it partonic} moments $L_{ij}$ computed with the 
lepton energy cut of $1~{\rm GeV}$ and the hadronic energy cut 
$M_H < 2.5~{\rm GeV}$. This set of cuts was previously studied in 
Ref.~\cite{ural1}.

\begin{tiny}
\begin{table*}[t]
\vspace{0.1cm}
\begin{center}
\begin{tabular}{|c|c|c|c|c|c|}
\hline\hline
$i$ & j & $L_{ij}^{(0)}$ & $L_{ij}^{(1)}$ & $L_{ij}^{(2,{\rm BLM})}$ & 
$L_{ij}^{(2)}$ \\ \hline\hline
$0$ & $0$  & 0.87135  &  -2.261(4) &  -27.7(1)  & 5.1(1) \\ \hline
$0$ & $1$  & 0.29306  &  -0.738(2) &  -8.13(1) & 1.38(2) \\ \hline 
$0$ & $2$  & 0.10789  &  -0.2558(8) &  -2.55(1)  & 0.38(1) \\ \hline 
$0$ & $3$  & 0.04210  &  -0.0920(4) &  -0.815(2) & 0.091(6) \\ \hline
$1$ & $0$  & 0.0  &  0.13110(7) & 2.231(1)  &  -0.638(3) \\ \hline
$1$ & $1$  & 0.0  &  0.05265(3) & 0.882(1)  &  -0.256(1) \\ \hline 
$1$ & $2$  & 0.0  &  0.02207(2) & 0.365(1)  &  -0.106(1)\\ \hline 
$2$ & $0$  & 0.0  &  4.973(4) $\cdot 10^{-3}$ & 6.83(1) $\cdot 10^{-2}$  & -9.8(1) $\cdot 10^{-3}$ \\ \hline 
$2$ & $1$  & 0.0  &  2.144(2) $\cdot 10^{-3}$ & 2.93(1)  $\cdot 10^{-2}$ & -4.3(1) $\cdot 10^{-3}$ \\ \hline 
$3$ & $0$  & 0.0  &  3.452(7) $\cdot 10^{-4}$ & 4.41(1)  $\cdot 10^{-3}$ & -4.9(1) $\cdot 10^{-4}$ \\ \hline \hline
\end{tabular}
\caption{\label{table1}  Moments of the partonic invariant mass $M_X^2$ and 
the partonic energy $E_X$ with the hadronic invariant mass cut 
 $M_{H} < 2.5~{\rm GeV}$ and the charged lepton energy 
cut $E_l < 1~{\rm GeV}$. See text for details. 
}
\vspace{-0.1cm}
\end{center}
\end{table*}
\end{tiny}

It follows from Table~\ref{table1} that similar to the total rate BLM and 
non-BLM corrections have opposite size, so that the total result for NNLO 
corrections is {\it smaller}  than the BLM corrections taken alone. 
The non-BLM corrections seem to be more important for lower-moments than 
for higher moments. Indeed, the ratio $L_{ij}^{(2)}/L_{ij}^{2,\rm BLM}$ 
decreases 
monotonically by about a factor of $1.6$,  
from $0.1842$ to $0.112$, for $i=0$ and $j$ changing from $j=0$ to $j=3$.
This trend is also visible in the absolute magnitude of the corrections.
Taking $\alpha_s(m_b) = 0.24$, we find that for $i=0,j=0$, 
the non-BLM corrections increase the moment  by about three percent while 
for $i=0,j=3$, they become as small as one percent. 

While these corrections look small compared to the current 
${\cal O}(10\%)$ uncertainty in the 
$|V_{ub}|$ determined from inclusive 
decays, we note that Ref.~\cite{ural2} estimates 
the   total theoretical uncertainty 
on $|V_{ub}|$ that can be achieved with  
various kinematic  cuts  on $M_H,E_e$, and $q^2$ to be 
close  to six  percent. 
The uncertainty in perturbative corrections, which mainly refers 
to non-BLM ${\cal O}(\alpha_s^2)$ effects that we discuss 
in this paper, is believed \cite{ural2} to be 
responsible for $30$ to $50\%$  of the 
full theory uncertainty. Our calculation allows to remove this part 
of the theory 
uncertainty by providing explicit results for non-BLM corrections. 

For example, one of the scenarios considered in Ref.~\cite{ural2} 
is  a high-cut on the lepton energy $E_l > 2~{\rm GeV}$; it corresponds 
to the measurement by the BABAR collaboration reported in Ref.~\cite{babar2}.
We find the non-BLM correction to $L_{00}(E_l > 2~{\rm GeV})$ 
using  the cumulative histogram in 
Fig.~\ref{fig2} and observe that 
it changes the leading order moment 
$L_{00}^{(0)}(E_l > 2~{\rm GeV}) = 0.257$  
by $6\%$.\footnote{For comparison, 
we note that the corresponding BLM corrections 
to  $L_{00}^{(0)}$  is $-30\%$.}    
Since the  experimental measurement corresponds 
to $|V_{ub}|^2 L_{00}$,  a $6\%$  shift in $L_{00}$ 
due to non-BLM corrections
translates  into a  $-3 \%$ shift  in $V_{ub}$.  We stress that 
the above number is given to illustrate the  magnitude of the 
expected effect; a precise statement about the impact of non-BLM 
corrections requires a dedicated analysis along the lines 
of Ref.~\cite{ural2}.  However, it is clear that 
our computation  should help in removing  a 
significant fraction of the  full theory error 
in $|V_{ub}|$ as estimated in \cite{ural2} for the  
$E_l > 2~{\rm GeV}$ cut.

We also note that it is customary to consider 
{\it normalized}  moments, which are defined as  
$C_{ij} = L_{ij}/L_{00}$. Since  both the numerator and the denominator 
in the definition of $C_{ij}$ receive perturbative corrections, 
we need to consistently expand $C_{ij}$ in a series in $\alpha_s$ 
to establish how stable it is against radiative corrections. 
 We find that  in case of $C_{0j}$, the non-BLM corrections 
are close to one-fifth of the BLM corrections  for all values 
of $j$ and they change the normalized 
moment by $-0.5\%$ for $j=1$ and by  $-1.45\%$ for $j=3$.

The situation changes 
dramatically for partonic invariant mass moments 
$L_{ij}$,  with 
$i \ne 0$. In this case the leading order partonic moments vanish since 
in the $b \to u e^- \bar \nu$ process the partonic invariant mass is zero.
As the result, for these moments our NNLO calculation is, essentially, 
next-to-leading order and the significance of non-BLM corrections increases. 
As follows from Table~\ref{table1} for 
 $L_{ij}$ moments with $i \ne 0$ and $j \ne 0$, the non-BLM 
corrections can be as large as  $30 \%$.

To conclude,  we presented  a computation of 
 ${\cal O}(\alpha_s^2)$  corrections  to  the fully-differential  decay rate 
of charmless semileptonic $b$ decay, 
$b \to X_u e \bar \nu$.  Our calculation provides a NNLO QCD 
description of 
arbitrary infra-red safe observables   and 
allows arbitrary kinematic cuts
including those that closely match the ones employed in experimental 
analyses.  We constructed a parton-level 
Monte-Carlo integrator which can be used  to compute 
large number of relevant observables 
and kinematic distributions in a single run of the program. 
This calculation, together 
with earlier results on NNLO QCD 
corrections to fully-differential 
$b \to c l \bar \nu$  transition \cite{Melnikov:2008qs,Biswas:2009rb},
makes all inclusive semileptonic 
decays of $b$-quarks upgraded to that  accuracy.
We hope that these results will contribute to the reduction of the 
theoretical error  on  $|V_{ub}|$ and $|V_{cb}|$  that will be achieved
in the forthcoming  $B$-physics experiments.

\vspace*{0.2cm}

{\bf Acknowledgments}
We are grateful to P.~Gambino and N.~Uraltsev for comments 
on the  manuscript. 
The research of F.C. and K.M.  is supported by US NSF under grant PHY-1214000.
The   research of K.M. and M.B. 
is partially 
supported by Karlsruhe Institute of Technology through a grant provided  
by its Distinguished Researcher Fellowship program.
M.B. is partially supported by the DFG through 
the SFB/TR~9 ``Computational particle physics''. 
Calculations described in this paper were performed at the Homewood 
High Performance Computer Cluster at Johns Hopkins University.

\end{document}